\documentclass[aps,prl,amssymb,showpacs,amsmath,reprint]{revtex4-2}
\usepackage{graphicx}
\usepackage{graphicx}
\usepackage{amsmath}
\usepackage{amsfonts}
\usepackage{xcolor}
\usepackage[T2A]{fontenc}
\providecommand{\LyX}{L\kern-.1667em\lower.25em\hbox{Y}\kern-.125emX\@}

\begin{document}
\title{Cyclops states in repulsive Kuramoto networks: the role of higher-order coupling}



\author{Vyacheslav O. Munyayev$^{1}$, Maxim I. Bolotov$^{1}$, Lev A. Smirnov$^{1}$, Grigory V. Osipov$^{1}$, and Igor Belykh$^{2}$\footnote{Corresponding author, e-mail: ibelykh@gsu.edu}}

\address{$^1$Department of Control Theory, Lobachevsky State University of Nizhny Novgorod,
	23 Gagarin Avenue, Nizhny Novgorod, 603022, Russia\\
	$^2$Department of Mathematics and Statistics and Neuroscience Institute, Georgia State University, P.O. Box 4110, Atlanta, Georgia, 30302-410, USA}	
\begin{abstract}
Repulsive oscillator networks can exhibit multiple cooperative rhythms, including chimera and cluster splay states. Yet, understanding which rhythm prevails remains challenging. Here, we address this fundamental question in the context of Kuramoto--Sakaguchi networks of identical rotators with higher-order coupling.  Through analysis and numerics, we show that three-cluster splay states with two distinct coherent clusters and a solitary oscillator are the prevalent rhythms in networks with an odd number of units. We denote such tripod patterns cyclops states with the solitary oscillator reminiscent of the Cyclops's eye. As their mythological counterparts, the cyclops states are giants that dominate the system's phase space in weakly repulsive networks with first-order coupling. Astonishingly, the addition of the second or third harmonics to the Kuramoto coupling function makes the cyclops states global attractors practically across the full range of coupling's repulsion. At a more general level, our results suggest clues for finding dominant rhythms in repulsive physical and biological networks. 
\end{abstract}

\date{\today}
\draft \maketitle

\textit{Introduction.} 
Networks of phase oscillators have been widely used as a paradigmatic model for emergent collective dynamics in  real-world systems, including neuronal networks \cite{hoppensteadt2012weakly}, populations of chemical oscillators \cite{tinsley2012chimera}, and power grids \cite{motter2013spontaneous,dorfler2013synchronization}. The Kuramoto model of one-dimensional (1D) \cite{kuramoto1975self, strogatz2000kuramoto} or two-dimensional (2D) phase oscillators \cite{ermentrout} is a prime example of such networks that can exhibit extraordinary collective dynamics \cite{acebron,barreto2008synchronization, ott2008low,hong2007entrainment, pikovsky2008partially, maistrenko2004mechanism,dorfler2011critical}, including full \cite{tanaka1997first,tanaka1997self,ji2014low,munyaev2020analytical,komarov2014synchronization}, partial  \cite{martens2009exact,barabash2021partial}, explosive \cite{gomez2011explosive,ji2013cluster,skardal2014disorder} and asymmetry-induced  synchronization \cite{nishikawa2016symmetric,nicolaou2019multifaceted}, 
 chimeras  \cite{kuramoto2002coexistence, abrams2004chimera, abrams2008solvable, panaggio2015chimera, zakharova2014chimera,panaggio2015chimera,bolotov2016marginal,bolotov2018simple}, solitary states  \cite{jaros2015chimera,maistrenko2017smallest,jaros2018solitary,teichmann2019solitary,munyayev2022stability}, clusters \cite{belykh2016bistability,brister2020three,ronge2021splay}, and generalized splay states \cite{berner2021generalized}. Notably, full synchronization is the most probable outcome and dominant rhythm induced by increasing all-to-all coupling in the classical Kuramoto model. Splay \cite{tsimring2005repulsive,gao2019repulsive}, generalized and cluster splay states \cite{berner2021generalized,ronge2021splay} are typically observed in Kuramoto networks with repulsive coupling; however, there is no complete understanding under which conditions a particular rhythm can emerge and become dominant.  Evidently, two repulsively coupled oscillators have a tendency to achieve anti-phase synchronization; however, predicting an outcome of such interactions in large repulsive networks is often elusive.  In particular, such interactions can lead to counterintuitive effects \cite{belykh2008weak,nishikawa2010network,belykh2015synergistic,reimbayev2017two}.

Equally important for relating Kuramoto networks to realistic physical systems is to understand the role of higher-order coupling terms which represent a Fourier decomposition of a general $2\pi$-periodic interaction function \cite{delabays2019dynamical}. Examples in which higher-order terms play a significant role include generalized Kuramoto-type models of neuronal plasticity and Hebbian learning \cite{seliger2002plasticity,niyogi2009learning}, coupled electrochemical oscillators \cite{kiss2005predicting}, and Josephson junctions \cite{goldobin2013phase}. It was previously
shown that the addition of higher-order terms to the classical Kuramoto model of 1D oscillators with all-to-all attractive coupling can induce a multiplicity of synchronous states \cite{komarov2013multiplicity} and switching between clusters of synchrony \cite{skardal2011cluster}. However, the role of higher-order coupling in rhythmogenesis in repulsive networks remains to be explored.

\begin{figure}[t]
	\includegraphics[width=0.85\columnwidth]{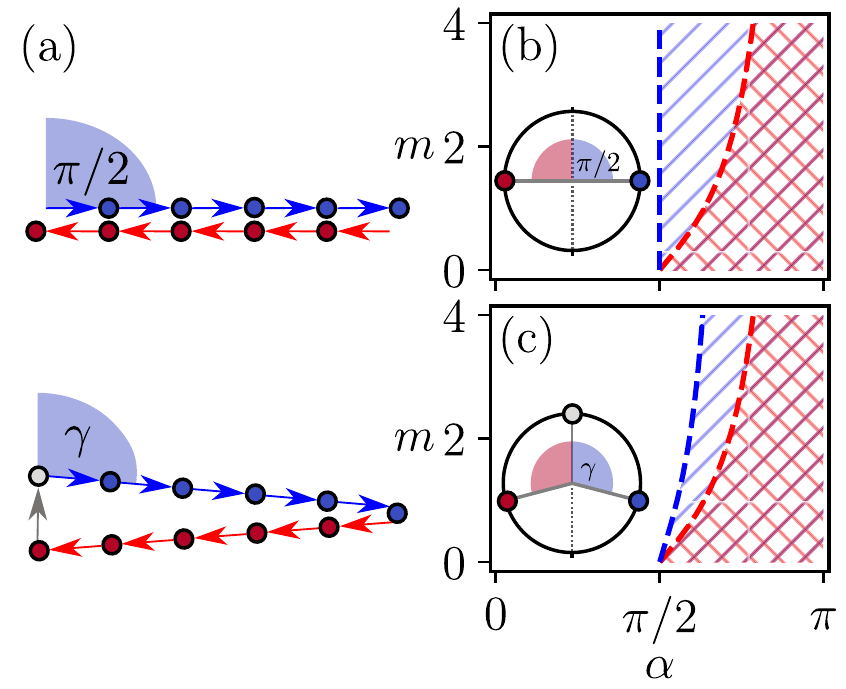}
	\caption{(a) Generalized splay state with a maximum $r_2$ in networks with even and odd $N$: a two-cluster splay state for $N=10$ (top) and a three-cluster, cyclops state for $N=11.$ The angle of the incoming arrow indicates the oscillator's phases; oscillators depicted by the same color have the same phase. The unit length of each arrow corresponds to $|z_k|=1$. (b) Local stability diagram $(\alpha,m)$ for the two-cluster splay state with $r_2=1.$ The blue (red) hatched area corresponds to the stability condition  \eqref{eq:stability_criterion} for two-cluster splay state with $r_2=1$ (generalized splay states with $r_2=0$) with the blue (red) dashed line plotting the equality condition in \eqref{eq:stability_criterion} with $r_2=1$ ($r_2=0$). The double hatched area is the region of stable co-existence of all generalized splay states with $0<r_2<1$. (c) Diagram similar to (b) but calculated for the three-cluster, cyclops state with $r_2=(N-3)/(N-1).$  The circles show the phase distributions $\theta_k$ for the two-cluster state (b) and for the cyclops state (c). Phase angle $\gamma=\arccos\bigl(1\big/(1-N)\bigr)$.
	\vspace{-5mm}	
	}
	\label{fig1}
\end{figure}

In this Letter, we make essential steps towards solving this critical problem for repulsive Kuramoto--Sakaguchi networks of identical 2D phase oscillators with phase-lagged first-order and higher-order coupling. We first show that two-cluster and three-cluster splay states are the dominant rhythms in weakly repulsive networks of even and odd numbers of oscillators with first-order coupling, respectively.  The three-cluster splay states are formed by two distinct coherent clusters and a solitary oscillator. These tripod states may be viewed as a hybrid that unites a two-body chimera with a solitary state. Inspired by the imposing single-eyed giant of Greek mythology, we call these tripod patterns cyclops states with the solitary oscillator and synchronous clusters representing the Cyclops's eye and shoulders, respectively. We report a surprising find that the addition of higher-order coupling terms induces global stability of cyclops states in practically the entire range of the phase-lag parameter that controls repulsion.\\
\textit{The network model.} We consider the Kuramoto--Sakaguchi network of 2D phase oscillators 
\begin{equation}
	m\ddot\theta_j+\dot\theta_j=\omega+\sum\limits_{n=1}^N \sum\limits_{q=1}^{l}\frac{K_q}{N}
	{\sin\left[q\left(\theta_{k}-\theta_j\right)-\alpha_q\right]},	\label{eq:system_origin}
\end{equation}
where variables $\theta_j\equiv \theta_i \,({\rm mod}\, 2\pi),$ $j=1,...,N$  are the oscillators' phases and the $l$th-order Kuramoto--Sakaguchi coupling \cite{sakaguchi2006instability} represents a pairwise interaction function $H(\theta_j,\theta_k).$ The oscillators are assumed to be identical, with frequency $\omega,$ inertia $m,$ and phase lags $\alpha_q \in [0, \pi).$ We set the coupling $K_1=1$ and phase lag $\alpha_1=\alpha.$\\
I. {\it First-order coupling: $l=1$.} In this simplest case, 
the system \eqref{eq:system_origin} can be cast into the form \cite{acebron}:
\begin{equation}
	\begin{gathered}
		m\ddot\theta_j\!+\!\dot\theta_{j}=\omega\!+\!\operatorname{Im}\!\left[R_1\!\left(t\right)e^{-i\left(\theta_j+\alpha\right)}\right]\!,\\
		R_1\!\left(t\right)=\frac{1}{N}\!\sum\limits_{k=1}^N\!{e^{i\theta_{k}}}=r_1 e^{i\psi_1},
	\end{gathered}
	\label{system}
\end{equation}
where $r_1$ and $\psi_1$ define the magnitude and the phase of
the first moment of the Kuramoto order parameter $R_1(t)$, respectively. The scalar $r_1$ characterizes the degree of phase synchrony. The synchronous solution $D(1)=\{\theta_1=...=\theta_N\}$ with $r_1=1$ is unstable for $\alpha \in (\pi/2, \pi)$ due to repulsive coupling \cite{munyayev2022stability,ronge2021splay}. Instead, the system \eqref{system} with $\alpha \in (\pi/2, \pi)$ is known to exhibit stable generalized splay states with a non-uniform phase distribution \cite{berner2021generalized} for intermediate values of inertia $m$ and rotatory solitary states \cite{munyayev2022stability} for larger $m$ which promotes rotatory dynamics \cite{belykh2016bistability}. In the following, we limit our attention to intermediate $m$ and analyze the prevalence of generalized splay states which represent phase-locked solutions 
$\theta_j=\omega t+\varphi_j,$ $j=1,...,N$ with constant relative phases $\varphi_j\in [0,\;2\pi]$ which satisfy the condition $R_1\!\left(t\right)=0$. The degree of cluster synchrony within a given splay state is controlled by 
the second moment of the Kuramoto order parameter, $	R_2\!\left(t\right)=N^{-1}\!\sum\limits_{k=1}^N\!{e^{i 2\theta_{k}}}=r_2 e^{i\psi_2}$ \cite{skardal2011cluster,berner2021generalized}. Remarkably, $r_2$ also controls the stability of the generalized splay state. Our stability analysis shows that  a cluster splay state with a given $r_2$ is locally stable in the parameter region:
\begin{equation}
	\cos\alpha<\frac{1}{m}-\sqrt{\frac{1}{m^2}+1-r_2^2}.
	\label{eq:stability_criterion}
\end{equation}
Although derived using a different argument, the condition \eqref{eq:stability_criterion} is similar to Corollary~9 in the previous stability study \cite{berner2021generalized}. Note that the right-hand side of inequality \eqref{eq:stability_criterion} is always non-positive thereby suggesting that generalized splay states can only be stable in the range of repulsive coupling which yields negative values of $\cos \alpha.$ The condition  \eqref{eq:stability_criterion} also suggests that increasing the degree of cluster synchrony $r_2$ enlarges the parameter region $(\alpha,m)$ for the stability of generalized splay states. The size of this region is maximized for generalized splay states with a maximum $r_2.$ As for 1D Kuramoto phase oscillators  \cite{skardal2011cluster,ronge2021splay}, the maximum value $r_2=1$ for generalized splay states with $r_1=0$ in the network \eqref{eq:system_origin} with even $N$ yields a two-cluster symmetric state: $\varphi_1=\ldots=\varphi_{N/2}=0$ and $\varphi_{N/2+1}=\ldots=\varphi_{N}=\pi$ with a relative phase angle of $\pi/2$ (Fig.~\ref{fig1}a). In accordance with  \eqref{eq:stability_criterion}, the two-cluster splay state is locally stable for any $\alpha \in (\pi/2,\pi)$ and any value of inertia $m>0$ (Fig.~\ref{fig1}b).

Finding generalized splay states which yield maximum values of $r_2$ for the network \eqref{system} with odd $N$ is more challenging. This problem amounts to finding the global maximum of $r_2=\text{Re} R_2$ subject to $R_1=0$ and $\text{Im} R_2=0$. We solve this optimization problem by the method of Lagrange multipliers via constructing the Lagrange function
\begin{equation}
\begin{array}{l}
	L=\text{Re} R_2\!\!-\!\!\lambda_1 \text{Re} R_1\!\!-\!\!\lambda_2 \text{Im} R_1\!\!-\!\!\lambda_3 \text{Im} R_2 =\\
	=\!N^{-1}\!\!\sum\limits_{k=1}^N\!\left(\cos{2{\theta_k}}\!\!-\!\!\lambda_1\cos{\theta_k}\!\!-\!\!\lambda_2\sin{\theta_k}\!\!-\!\!\lambda_3\sin{2{\theta_k}}\right)\!,
	\end{array}
	\label{eq:function_lagrange}
\end{equation}
where $\lambda_1$, $\lambda_2$ and $\lambda_3$ are scalars (multipliers). Solving $\nabla_{\theta_1,...,\theta_N,\lambda_1,\lambda_2,\lambda_3} L=0$ yields the necessary conditions for finding local extrema of $\text{Re} R_2:$
\begin{equation}
	\begin{array}{c}
		\sum\limits_{k=1}^N(-2\sin{2{\theta_k}}+\lambda_1\sin{\theta_k}-\lambda_2\cos{\theta_k}-2\lambda_3\cos{2{\theta_k}})=0,\\
		R_1=0,\quad
	\text{Im} R_2=0. \label{eq:equations_lagrange}
	\end{array}
\end{equation}
For the given side conditions, we obtain $\lambda_3=0.$ Further analysis of  \eqref{eq:equations_lagrange} can be simplified by introducing complex variables $z_k=e^{i\theta_k}$ and turning \eqref{eq:equations_lagrange} into
\begin{equation}
	\begin{gathered}
		z^4-\lambda z^3+\lambda^{*}{z}-1=0, \quad
		\left|z\right|=1,
	\end{gathered}
	\label{eq:equations_lagrange_a_complex}
\end{equation}
where the subscript $k$ has been omitted for brevity and $\lambda=\left(\lambda_1-i\lambda_2\right)/2$. To satisfy the 
condition $R_1=0$, the fourth-order equation \eqref{eq:equations_lagrange_a_complex} must have at least three distinct roots $\xi_1$, $\xi_2$, $\xi_3$. As a result, three- or four-cluster splay states correspond to local extrema of $r_2$ such that for odd $N,$
the cluster partition that maximizes $r_2$ is
\begin{equation}
	\begin{gathered}
		N_{1}\xi_{1}+N_{2}\xi_{2}+N_{3}\xi_{3}+N_{4}\xi_{4}=0,\quad
	\left|\xi_{1,2,3,4}\right|=1, 
	\end{gathered}
	\label{eq:partition}
\end{equation}
where $N_p$ and $\xi_p=z_p,$ $p=1,..,4$ are the size and complex phase of the $p$th cluster, respectively. Here, 
$N_4$ may be equal to $0$ in the case of a three-cluster state. In geometrical terms, finding an algebraic partition satisfying to \eqref{eq:partition} is analogous to finding all possible quadrilaterals (triangles for the three-cluster states) with the perimeter $N$ and integer side lengths (see Fig.~\ref{fig1}a). Performing such an exhaustive search for odd $N$, we conclude that four-cluster partitions can only yield a local maximum $r_2=(N-3)/N$ which is reached at the four-cluster splay state:
$\varphi_{1}=\ldots=\varphi_{\left(N-3\right)/2}=0$, $\varphi_{\left(N-1\right)/2}=\pi/3$, $\varphi_{\left(N+1\right)/2}=-\pi/3$, $\varphi_{\left(N+3\right)/2}=\ldots=\varphi_{N}=\pi$ subject to an arbitrary constant phase shift. The global 
maximum of $r_2=(N-3)/(N-1)$ is reached at a continuum of three-cluster splay states
\begin{equation}
	\begin{array}{l}
\varphi_1=\varphi_{2}=\ldots=\varphi_{\left(N-1\right)/2}=\gamma, \\
\varphi_{\left(N-1\right)/2+1}=\ldots=\varphi_{N-1}=-\gamma \quad \mbox{\rm and}\quad \varphi_{N}=0,
\end{array}
\label{cyclope}
\end{equation}
where $\gamma=\arccos\bigl(1\big/(1-N)\bigr)$ and the choice of the reference zero phase for $\varphi_N$ is arbitrary.
The expression for $\gamma$ can be verified from the triangle in Fig.~\ref{fig1}a such that $\cos \gamma=-\cos(\pi-\gamma)=1/(1-N).$ The calculation of the global maximum $r_2$ for the three-cluster state \eqref{cyclope}
can be performed via 
$$R_2\!\left(t\right)=\frac{1}{N}\!\sum\limits_{k=1}^N\!{e^{i 2\theta_{k}}}=\frac{N-1}{2N} e^{2 i\gamma}+
\frac{N-1}{2N} e^{-2 i\gamma}+\frac{1}{N}$$ 
which yields $r_2= \text{Re} R_2=N^{-1}(N-1)\cos 2\gamma+N^{-1}=(N-3)/(N-1)$ due to $\cos 2\gamma=2\cos^2 \gamma-1=2{(1-N)^{-2}}-1.$

\begin{figure}[h!]
	\includegraphics[width=0.9\columnwidth]{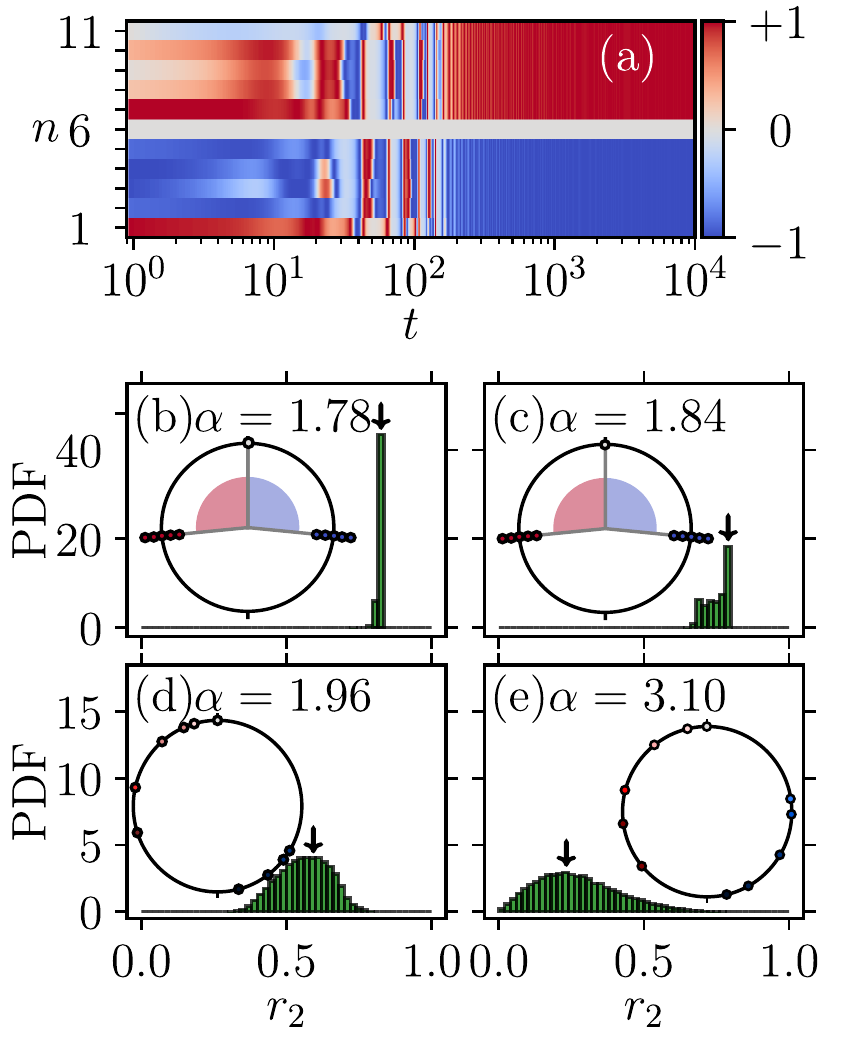}
	\caption{(a) The onset of a symmetric cyclops state from randomly chosen initial conditions. The colors depict $\sin{(\theta_n(t) - \theta_{6}(t))},$ where the $6th$ element is a solitary oscillator. (b-e).
		Histograms for a numerically calculated probability density function (PDF) of the $r_2$ distribution for the established rhythms. Remarkably, all these rhythms are generalized splay states with $r_1 = 0$. The number of trials: 
		$50,\!000$ from randomly generated initial conditions for $\theta_n$ and $\dot{\theta}_n,$ $n=1,...,11.$ The PDF is normalized over $50$ bins. The circles show the phase distributions $\theta_j$ for the most probable $r_2$ (indicated by the arrow above the bins). (b) The dominant cyclops state from (a) with $\alpha=1.78.$ (c) $\alpha=1.84.$ (d) $\alpha=1.96.$ (e) $\alpha=3.10.$ Other parameters: $N=11$, $m=1.0$ and $\omega=1.0$.
		\vspace{-5mm}		
	}
	\label{fig2}
\end{figure}
Thus, out of all possible generalized splay states in the networks with odd $N,$ the three-cluster splay state \eqref{cyclope} has the largest local stability region in the system's parameter space and therefore is most abundant. The three-cluster splay state  has a distinct structure composed of two equally sized clusters symmetric about a solitary oscillator, reminiscent of the Cyclops's eye.  In Greek mythology, the Cyclopes were one-eyed giants who were famed for their ability to build impressive structures. This is also relevant to the three-cluster splay states \eqref{cyclope} that, as we will see, can make up an impressive skeleton of dominant states in the system's phase space. Given their shapes and possible prevalence, we call them symmetric cyclops states. Generalizing this concept to  three-cluster states  \eqref{cyclope} with an asymmetry in the phases of the  synchronous clusters relative to the solitary oscillator, we will term them asymmetric cyclops states.

\begin{figure}[h]\center
	\includegraphics[width=0.85\columnwidth]{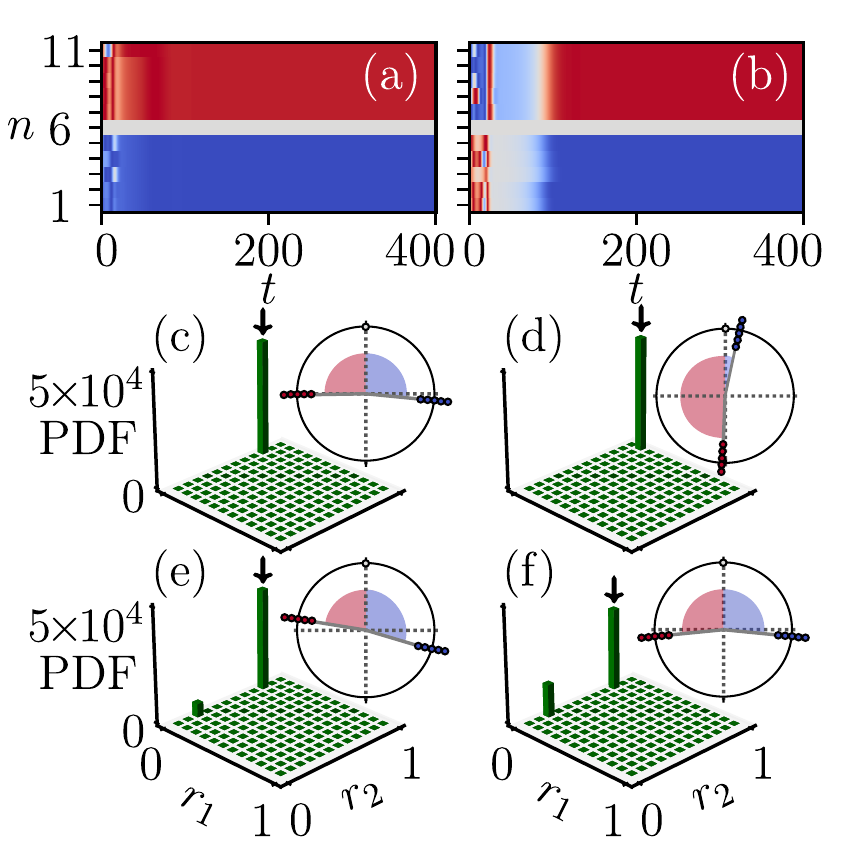}
	\caption{The role of the second (a, c, d) and third (b, e, f) coupling harmonics. 
		Histograms for a numerically calculated PDF of the $r_1,r_2$ distribution for the established rhythms. The number of trials: $50,\!000$ from randomly generated initial conditions. Symmetric and asymmetric cyclops states become the dominant rhythms in both cases of weak repulsive ((a),(c) and (b),(e) with $\alpha = 1.96$) and strong repulsive coupling ((d),(f) with $\alpha = 3.10$). Time series (a) and (b) correspond to the cyclops states in (c) and (e), respectively. Other parameters: $N=11$, $m=1.0$, $\omega=1.7$ and $K_2=0.05$, $\alpha_2=0.3$, $K_3=0.0$, $\alpha_3=0.0$ (c,d); $K_2=0.05$, $\alpha_2=0.3$, $K_3=0.1$, $\alpha_3=1$ (e,f).
	\vspace{-5mm}	
	} \label{fig3}
\end{figure}

\begin{figure}[h]\center
	\includegraphics[width=0.75\columnwidth]{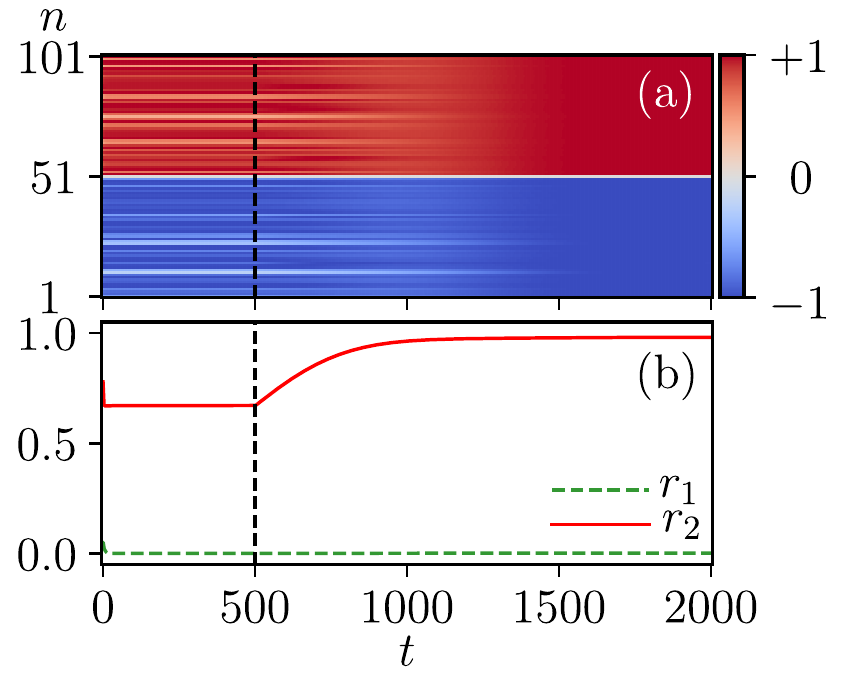}
	\caption{The role of the second harmonic in stabilizing a cyclops state in  system  \eqref{eq:system_origin} with $N=101$, $m=1.0$, $\omega=1.7$, $\alpha=3.1$. 
	The system with only the first-order coupling ($K_2=0$) evolves into a generalized splay state with $r_1=0$ from random initial conditions for  $0<t<500$. Switching on the second harmonic with $K_2=0.002$ and $\alpha_2=0.2$ induces a stable cyclops state ($500<t<2000$). (a) Colors indicate $\sin{(\theta_n(t)\!-\!\theta_{51}(t))}$. (b) The corresponding values of $r_1$ and $r_2$.
	\vspace{-5mm}
	}	\label{fig4}
\end{figure}

Figure~\ref{fig2} shows that symmetric cyclops states are the dominant states in the network with $N=11$ for the values of $\alpha$ that represent weak repulsion. Increasing $\alpha$ makes other generalized splay states with lower $r_2$ more prevalent. Similar effects are observed in larger-size networks.
Our numerical analysis of the prevalence of two-cluster states with the maximum $r_2=1$  in the network with $N=10$ yields a diagram quite similar to Fig.~\ref{fig2} and therefore not shown. This indicates that the two-cluster state is also dominant in weakly repulsive networks with even $N.$\\
II. {\it Higher-order coupling: $l=2$ and $l=3.$} The addition of the second-order ($l=2$) and also the third-order coupling ($l=3$) to the first-order coupling network preserves the existence of the two-cluster and cyclops states. Astonishingly, this addition also makes the cyclops states global attractors practically across the full range of phase lag $\alpha \in (\pi/2,\pi)$ corresponding to repulsive coupling (Fig.~\ref{fig3}). Note that the repulsiveness of the higher-order coupling in the system \eqref{eq:system_origin} is controlled by a combination of phase lags $\alpha,\alpha_2,\alpha_3$ and coupling strengths $K_2,K_3$ via the condition $H'(\theta_k,\theta_k)<0$ that guarantees the instability of the synchronous solution $D(1).$ The particular choices of $\alpha_2,\alpha_3$ and $K_2,K_3$ used in Fig.~\ref{fig3} preserve this repulsiveness for any $\alpha \in (\pi/2,\pi).$ In the case of strongly repulsive coupling $\alpha$ close to $\pi,$ the second-order harmonics induces the prevalent asymmetric cyclops states (Fig.~\ref{fig3}d) while the addition of the third-harmonics makes these cyclops states symmetric (Fig.~\ref{fig3}f). 
The higher-harmonics also have the same stabilization effect on the dominance of the  two-cluster splay states in networks with even $N$.

A detailed analysis of the decisive role of the higher-order coupling in the 
appearance and prevalence of the cyclops states will be reported in a more technical publication. In simple terms, the effect can be understood via a closer inspection of the coupling function $H(x)=\sin(x-\alpha)+K_2 \sin(2x-\alpha_2)+K_3 \sin(3x-\alpha_3),$ where $x=\theta_k-\theta_j.$ Here, the double and triple angle phase difference single out the ranges of $x$ in which the second or third harmonics play a role of attractive coupling when $-\pi/2<2x-\alpha_2<\pi/2$ or $-\pi/2<3x-\alpha_3<\pi/2.$ As a result, these harmonics can promote the formation of two synchronous clusters of oscillators with the phases that fall into the select ranges of $x.$ At the same time, the first repulsive harmonics maintains the balance among the clusters and the solitary oscillator.

This effect is also present in large networks. Figure~\ref{fig4} provides evidence that the activation of the second-order harmonics turns a generalized splay state of the network with $N=101$ into a cyclops state even if the repulsion is strong ($\alpha$ close to $\pi$).
Our preliminary studies show that the prevalence of cyclops states persists in the presence of small 
intrinsic frequency mismatch and noise.\\
\textit{Conclusions.} In this work, we studied Kuramoto--Sakaguchi networks of identical 2D rotators  to reveal the surprising role of higher-order harmonics in inducing stable two-cluster and cyclops splay states as a result of a complex interplay between the network size, inertia, and the phase lags. We offered compelling evidence in favor of the prevalence of these states in repulsive networks whose interactions could be modeled via higher-order harmonics of the Fourier decomposition of a coupling function. Beyond the globally coupled Kuramoto oscillators, we hypothesize that cyclops states could be dominant in densely coupled networks. Our results suggest that cyclops states may be viewed as a structural foundation for understanding and predicting emergent repulsive network dynamics in physical and biological networks, similar to the fundamental concept of full synchronization in attractive networks.  In a broader context, our  study leverages the role of high-order harmonics in stabilizing low-dimensional dynamical patterns in oscillatory networks.

\begin{acknowledgments}
This work was supported by the MSHE under project No. 0729-2020-0036 (to M.I.B), 
the RSF under project 22-12-00348 (to V.O.M., L.A.S. and G.V.O), and the NSF (USA) under grants DMS-1909924 and CMMI-2009329 (to I.B.). We thank V. Kostin for useful discussions.

\end{acknowledgments}


\begin{thebibliography}{56}%
	\makeatletter
	\providecommand \@ifxundefined [1]{%
		\@ifx{#1\undefined}
	}%
	\providecommand \@ifnum [1]{%
		\ifnum #1\expandafter \@firstoftwo
		\else \expandafter \@secondoftwo
		\fi
	}%
	\providecommand \@ifx [1]{%
		\ifx #1\expandafter \@firstoftwo
		\else \expandafter \@secondoftwo
		\fi
	}%
	\providecommand \natexlab [1]{#1}%
	\providecommand \enquote  [1]{``#1''}%
	\providecommand \bibnamefont  [1]{#1}%
	\providecommand \bibfnamefont [1]{#1}%
	\providecommand \citenamefont [1]{#1}%
	\providecommand \href@noop [0]{\@secondoftwo}%
	\providecommand \href [0]{\begingroup \@sanitize@url \@href}%
	\providecommand \@href[1]{\@@startlink{#1}\@@href}%
	\providecommand \@@href[1]{\endgroup#1\@@endlink}%
	\providecommand \@sanitize@url [0]{\catcode `\\12\catcode `\$12\catcode
		`\&12\catcode `\#12\catcode `\^12\catcode `\_12\catcode `\%12\relax}%
	\providecommand \@@startlink[1]{}%
	\providecommand \@@endlink[0]{}%
	\providecommand \url  [0]{\begingroup\@sanitize@url \@url }%
	\providecommand \@url [1]{\endgroup\@href {#1}{\urlprefix }}%
	\providecommand \urlprefix  [0]{URL }%
	\providecommand \Eprint [0]{\href }%
	\providecommand \doibase [0]{http://dx.doi.org/}%
	\providecommand \selectlanguage [0]{\@gobble}%
	\providecommand \bibinfo  [0]{\@secondoftwo}%
	\providecommand \bibfield  [0]{\@secondoftwo}%
	\providecommand \translation [1]{[#1]}%
	\providecommand \BibitemOpen [0]{}%
	\providecommand \bibitemStop [0]{}%
	\providecommand \bibitemNoStop [0]{.\EOS\space}%
	\providecommand \EOS [0]{\spacefactor3000\relax}%
	\providecommand \BibitemShut  [1]{\csname bibitem#1\endcsname}%
	\let\auto@bib@innerbib\@empty
	\bibitem [{\citenamefont {Hoppensteadt}\ and\ \citenamefont
		{Izhikevich}(2012)}]{hoppensteadt2012weakly}%
	\BibitemOpen
	\bibfield  {author} {\bibinfo {author} {\bibfnamefont {F.~C.}\ \bibnamefont
			{Hoppensteadt}}\ and\ \bibinfo {author} {\bibfnamefont {E.~M.}\ \bibnamefont
			{Izhikevich}},\ }\href@noop {} {\emph {\bibinfo {title} {Weakly connected
				neural networks}}},\ Vol.\ \bibinfo {volume} {126}\ (\bibinfo  {publisher}
	{Springer Science \& Business Media},\ \bibinfo {year} {2012})\BibitemShut
	{NoStop}%
	\bibitem [{\citenamefont {Tinsley}\ \emph {et~al.}(2012)\citenamefont
		{Tinsley}, \citenamefont {Nkomo},\ and\ \citenamefont
		{Showalter}}]{tinsley2012chimera}%
	\BibitemOpen
	\bibfield  {author} {\bibinfo {author} {\bibfnamefont {M.~R.}\ \bibnamefont
			{Tinsley}}, \bibinfo {author} {\bibfnamefont {S.}~\bibnamefont {Nkomo}}, \
		and\ \bibinfo {author} {\bibfnamefont {K.}~\bibnamefont {Showalter}},\
	}\href@noop {} {\bibfield  {journal} {\bibinfo  {journal} {Nature Physics}\
		}\textbf {\bibinfo {volume} {8}},\ \bibinfo {pages} {662} (\bibinfo {year}
		{2012})}\BibitemShut {NoStop}%
	\bibitem [{\citenamefont {Motter}\ \emph {et~al.}(2013)\citenamefont {Motter},
		\citenamefont {Myers}, \citenamefont {Anghel},\ and\ \citenamefont
		{Nishikawa}}]{motter2013spontaneous}%
	\BibitemOpen
	\bibfield  {author} {\bibinfo {author} {\bibfnamefont {A.~E.}\ \bibnamefont
			{Motter}}, \bibinfo {author} {\bibfnamefont {S.~A.}\ \bibnamefont {Myers}},
		\bibinfo {author} {\bibfnamefont {M.}~\bibnamefont {Anghel}}, \ and\ \bibinfo
		{author} {\bibfnamefont {T.}~\bibnamefont {Nishikawa}},\ }\href@noop {}
	{\bibfield  {journal} {\bibinfo  {journal} {Nature Physics}\ }\textbf
		{\bibinfo {volume} {9}},\ \bibinfo {pages} {191} (\bibinfo {year}
		{2013})}\BibitemShut {NoStop}%
	\bibitem [{\citenamefont {D{\"o}rfler}\ \emph {et~al.}(2013)\citenamefont
		{D{\"o}rfler}, \citenamefont {Chertkov},\ and\ \citenamefont
		{Bullo}}]{dorfler2013synchronization}%
	\BibitemOpen
	\bibfield  {author} {\bibinfo {author} {\bibfnamefont {F.}~\bibnamefont
			{D{\"o}rfler}}, \bibinfo {author} {\bibfnamefont {M.}~\bibnamefont
			{Chertkov}}, \ and\ \bibinfo {author} {\bibfnamefont {F.}~\bibnamefont
			{Bullo}},\ }\href@noop {} {\bibfield  {journal} {\bibinfo  {journal}
			{Proceedings of the National Academy of Sciences}\ }\textbf {\bibinfo
			{volume} {110}},\ \bibinfo {pages} {2005} (\bibinfo {year}
		{2013})}\BibitemShut {NoStop}%
	\bibitem [{\citenamefont {Kuramoto}(1975)}]{kuramoto1975self}%
	\BibitemOpen
	\bibfield  {author} {\bibinfo {author} {\bibfnamefont {Y.}~\bibnamefont
			{Kuramoto}},\ }in\ \href@noop {} {\emph {\bibinfo {booktitle} {International
				Symposium on Mathematical Problems in Theoretical Physics}}}\ (\bibinfo
	{organization} {Springer},\ \bibinfo {year} {1975})\ pp.\ \bibinfo {pages}
	{420--422}\BibitemShut {NoStop}%
	\bibitem [{\citenamefont {Strogatz}(2000)}]{strogatz2000kuramoto}%
	\BibitemOpen
	\bibfield  {author} {\bibinfo {author} {\bibfnamefont {S.~H.}\ \bibnamefont
			{Strogatz}},\ }\href@noop {} {\bibfield  {journal} {\bibinfo  {journal}
			{Physica D: Nonlinear Phenomena}\ }\textbf {\bibinfo {volume} {143}},\
		\bibinfo {pages} {1} (\bibinfo {year} {2000})}\BibitemShut {NoStop}%
	\bibitem [{\citenamefont {Ermentrout}(1997)}]{ermentrout}%
	\BibitemOpen
	\bibfield  {author} {\bibinfo {author} {\bibfnamefont {B.}~\bibnamefont
			{Ermentrout}},\ }\href@noop {} {\bibfield  {journal} {\bibinfo  {journal}
			{Journal of Mathematical Biology}\ } (\bibinfo {year} {1997})}\BibitemShut
	{NoStop}%
	\bibitem [{\citenamefont {Acebr{\'o}n}\ \emph {et~al.}(2005)\citenamefont
		{Acebr{\'o}n}, \citenamefont {Bonilla}, \citenamefont {Vicente},
		\citenamefont {Ritort},\ and\ \citenamefont {Spigler}}]{acebron}%
	\BibitemOpen
	\bibfield  {author} {\bibinfo {author} {\bibfnamefont {J.~A.}\ \bibnamefont
			{Acebr{\'o}n}}, \bibinfo {author} {\bibfnamefont {L.~L.}\ \bibnamefont
			{Bonilla}}, \bibinfo {author} {\bibfnamefont {C.~J.~P.}\ \bibnamefont
			{Vicente}}, \bibinfo {author} {\bibfnamefont {F.}~\bibnamefont {Ritort}}, \
		and\ \bibinfo {author} {\bibfnamefont {R.}~\bibnamefont {Spigler}},\
	}\href@noop {} {\bibfield  {journal} {\bibinfo  {journal} {Reviews of Modern
				Physics}\ }\textbf {\bibinfo {volume} {77}},\ \bibinfo {pages} {137}
		(\bibinfo {year} {2005})}\BibitemShut {NoStop}%
	\bibitem [{\citenamefont {Barreto}\ \emph {et~al.}(2008)\citenamefont
		{Barreto}, \citenamefont {Hunt}, \citenamefont {Ott},\ and\ \citenamefont
		{So}}]{barreto2008synchronization}%
	\BibitemOpen
	\bibfield  {author} {\bibinfo {author} {\bibfnamefont {E.}~\bibnamefont
			{Barreto}}, \bibinfo {author} {\bibfnamefont {B.}~\bibnamefont {Hunt}},
		\bibinfo {author} {\bibfnamefont {E.}~\bibnamefont {Ott}}, \ and\ \bibinfo
		{author} {\bibfnamefont {P.}~\bibnamefont {So}},\ }\href@noop {} {\bibfield
		{journal} {\bibinfo  {journal} {Physical Review E}\ }\textbf {\bibinfo
			{volume} {77}},\ \bibinfo {pages} {036107} (\bibinfo {year}
		{2008})}\BibitemShut {NoStop}%
	\bibitem [{\citenamefont {Ott}\ and\ \citenamefont
		{Antonsen}(2008)}]{ott2008low}%
	\BibitemOpen
	\bibfield  {author} {\bibinfo {author} {\bibfnamefont {E.}~\bibnamefont
			{Ott}}\ and\ \bibinfo {author} {\bibfnamefont {T.~M.}\ \bibnamefont
			{Antonsen}},\ }\href@noop {} {\bibfield  {journal} {\bibinfo  {journal}
			{Chaos: An Interdisciplinary Journal of Nonlinear Science}\ }\textbf
		{\bibinfo {volume} {18}},\ \bibinfo {pages} {037113} (\bibinfo {year}
		{2008})}\BibitemShut {NoStop}%
	\bibitem [{\citenamefont {Hong}\ \emph {et~al.}(2007)\citenamefont {Hong},
		\citenamefont {Chat{\'e}}, \citenamefont {Park},\ and\ \citenamefont
		{Tang}}]{hong2007entrainment}%
	\BibitemOpen
	\bibfield  {author} {\bibinfo {author} {\bibfnamefont {H.}~\bibnamefont
			{Hong}}, \bibinfo {author} {\bibfnamefont {H.}~\bibnamefont {Chat{\'e}}},
		\bibinfo {author} {\bibfnamefont {H.}~\bibnamefont {Park}}, \ and\ \bibinfo
		{author} {\bibfnamefont {L.-H.}\ \bibnamefont {Tang}},\ }\href@noop {}
	{\bibfield  {journal} {\bibinfo  {journal} {Physical Review Letters}\
		}\textbf {\bibinfo {volume} {99}},\ \bibinfo {pages} {184101} (\bibinfo
		{year} {2007})}\BibitemShut {NoStop}%
	\bibitem [{\citenamefont {Pikovsky}\ and\ \citenamefont
		{Rosenblum}(2008)}]{pikovsky2008partially}%
	\BibitemOpen
	\bibfield  {author} {\bibinfo {author} {\bibfnamefont {A.}~\bibnamefont
			{Pikovsky}}\ and\ \bibinfo {author} {\bibfnamefont {M.}~\bibnamefont
			{Rosenblum}},\ }\href@noop {} {\bibfield  {journal} {\bibinfo  {journal}
			{Physical Review Letters}\ }\textbf {\bibinfo {volume} {101}},\ \bibinfo
		{pages} {264103} (\bibinfo {year} {2008})}\BibitemShut {NoStop}%
	\bibitem [{\citenamefont {Maistrenko}\ \emph {et~al.}(2004)\citenamefont
		{Maistrenko}, \citenamefont {Popovych}, \citenamefont {Burylko},\ and\
		\citenamefont {Tass}}]{maistrenko2004mechanism}%
	\BibitemOpen
	\bibfield  {author} {\bibinfo {author} {\bibfnamefont {Y.}~\bibnamefont
			{Maistrenko}}, \bibinfo {author} {\bibfnamefont {O.}~\bibnamefont
			{Popovych}}, \bibinfo {author} {\bibfnamefont {O.}~\bibnamefont {Burylko}}, \
		and\ \bibinfo {author} {\bibfnamefont {P.}~\bibnamefont {Tass}},\ }\href@noop
	{} {\bibfield  {journal} {\bibinfo  {journal} {Physical Review Letters}\
		}\textbf {\bibinfo {volume} {93}},\ \bibinfo {pages} {084102} (\bibinfo
		{year} {2004})}\BibitemShut {NoStop}%
	\bibitem [{\citenamefont {D{\"o}rfler}\ and\ \citenamefont
		{Bullo}(2011)}]{dorfler2011critical}%
	\BibitemOpen
	\bibfield  {author} {\bibinfo {author} {\bibfnamefont {F.}~\bibnamefont
			{D{\"o}rfler}}\ and\ \bibinfo {author} {\bibfnamefont {F.}~\bibnamefont
			{Bullo}},\ }\href@noop {} {\bibfield  {journal} {\bibinfo  {journal} {SIAM
				Journal on Applied Dynamical Systems}\ }\textbf {\bibinfo {volume} {10}},\
		\bibinfo {pages} {1070} (\bibinfo {year} {2011})}\BibitemShut {NoStop}%
	\bibitem [{\citenamefont {Tanaka}\ \emph
		{et~al.}(1997{\natexlab{a}})\citenamefont {Tanaka}, \citenamefont
		{Lichtenberg},\ and\ \citenamefont {Oishi}}]{tanaka1997first}%
	\BibitemOpen
	\bibfield  {author} {\bibinfo {author} {\bibfnamefont {H.-A.}\ \bibnamefont
			{Tanaka}}, \bibinfo {author} {\bibfnamefont {A.~J.}\ \bibnamefont
			{Lichtenberg}}, \ and\ \bibinfo {author} {\bibfnamefont {S.}~\bibnamefont
			{Oishi}},\ }\href@noop {} {\bibfield  {journal} {\bibinfo  {journal}
			{Physical Review Letters}\ }\textbf {\bibinfo {volume} {78}},\ \bibinfo
		{pages} {2104} (\bibinfo {year} {1997}{\natexlab{a}})}\BibitemShut {NoStop}%
	\bibitem [{\citenamefont {Tanaka}\ \emph
		{et~al.}(1997{\natexlab{b}})\citenamefont {Tanaka}, \citenamefont
		{Lichtenberg},\ and\ \citenamefont {Oishi}}]{tanaka1997self}%
	\BibitemOpen
	\bibfield  {author} {\bibinfo {author} {\bibfnamefont {H.-A.}\ \bibnamefont
			{Tanaka}}, \bibinfo {author} {\bibfnamefont {A.~J.}\ \bibnamefont
			{Lichtenberg}}, \ and\ \bibinfo {author} {\bibfnamefont {S.}~\bibnamefont
			{Oishi}},\ }\href@noop {} {\bibfield  {journal} {\bibinfo  {journal} {Physica
				D: Nonlinear Phenomena}\ }\textbf {\bibinfo {volume} {100}},\ \bibinfo
		{pages} {279} (\bibinfo {year} {1997}{\natexlab{b}})}\BibitemShut {NoStop}%
	\bibitem [{\citenamefont {Ji}\ \emph {et~al.}(2014)\citenamefont {Ji},
		\citenamefont {Peron}, \citenamefont {Rodrigues},\ and\ \citenamefont
		{Kurths}}]{ji2014low}%
	\BibitemOpen
	\bibfield  {author} {\bibinfo {author} {\bibfnamefont {P.}~\bibnamefont
			{Ji}}, \bibinfo {author} {\bibfnamefont {T.~K.}\ \bibnamefont {Peron}},
		\bibinfo {author} {\bibfnamefont {F.~A.}\ \bibnamefont {Rodrigues}}, \ and\
		\bibinfo {author} {\bibfnamefont {J.}~\bibnamefont {Kurths}},\ }\href@noop {}
	{\bibfield  {journal} {\bibinfo  {journal} {Scientific Reports}\ }\textbf
		{\bibinfo {volume} {4}} (\bibinfo {year} {2014})}\BibitemShut {NoStop}%
	\bibitem [{\citenamefont {Munyaev}\ \emph {et~al.}(2020)\citenamefont
		{Munyaev}, \citenamefont {Smirnov}, \citenamefont {Kostin}, \citenamefont
		{Osipov},\ and\ \citenamefont {Pikovsky}}]{munyaev2020analytical}%
	\BibitemOpen
	\bibfield  {author} {\bibinfo {author} {\bibfnamefont {V.}~\bibnamefont
			{Munyaev}}, \bibinfo {author} {\bibfnamefont {L.}~\bibnamefont {Smirnov}},
		\bibinfo {author} {\bibfnamefont {V.}~\bibnamefont {Kostin}}, \bibinfo
		{author} {\bibfnamefont {G.}~\bibnamefont {Osipov}}, \ and\ \bibinfo {author}
		{\bibfnamefont {A.}~\bibnamefont {Pikovsky}},\ }\href@noop {} {\bibfield
		{journal} {\bibinfo  {journal} {New Journal of Physics}\ }\textbf {\bibinfo
			{volume} {22}},\ \bibinfo {pages} {023036} (\bibinfo {year}
		{2020})}\BibitemShut {NoStop}%
	\bibitem [{\citenamefont {Komarov}\ \emph {et~al.}(2014)\citenamefont
		{Komarov}, \citenamefont {Gupta},\ and\ \citenamefont
		{Pikovsky}}]{komarov2014synchronization}%
	\BibitemOpen
	\bibfield  {author} {\bibinfo {author} {\bibfnamefont {M.}~\bibnamefont
			{Komarov}}, \bibinfo {author} {\bibfnamefont {S.}~\bibnamefont {Gupta}}, \
		and\ \bibinfo {author} {\bibfnamefont {A.}~\bibnamefont {Pikovsky}},\
	}\href@noop {} {\bibfield  {journal} {\bibinfo  {journal} {EPL (Europhysics
				Letters)}\ }\textbf {\bibinfo {volume} {106}},\ \bibinfo {pages} {40003}
		(\bibinfo {year} {2014})}\BibitemShut {NoStop}%
	\bibitem [{\citenamefont {Martens}\ \emph {et~al.}(2009)\citenamefont
		{Martens}, \citenamefont {Barreto}, \citenamefont {Strogatz}, \citenamefont
		{Ott}, \citenamefont {So},\ and\ \citenamefont
		{Antonsen}}]{martens2009exact}%
	\BibitemOpen
	\bibfield  {author} {\bibinfo {author} {\bibfnamefont {E.~A.}\ \bibnamefont
			{Martens}}, \bibinfo {author} {\bibfnamefont {E.}~\bibnamefont {Barreto}},
		\bibinfo {author} {\bibfnamefont {S.}~\bibnamefont {Strogatz}}, \bibinfo
		{author} {\bibfnamefont {E.}~\bibnamefont {Ott}}, \bibinfo {author}
		{\bibfnamefont {P.}~\bibnamefont {So}}, \ and\ \bibinfo {author}
		{\bibfnamefont {T.}~\bibnamefont {Antonsen}},\ }\href@noop {} {\bibfield
		{journal} {\bibinfo  {journal} {Physical Review E}\ }\textbf {\bibinfo
			{volume} {79}},\ \bibinfo {pages} {026204} (\bibinfo {year}
		{2009})}\BibitemShut {NoStop}%
	\bibitem [{\citenamefont {Barabash}\ \emph {et~al.}(2021)\citenamefont
		{Barabash}, \citenamefont {Belykh}, \citenamefont {Osipov},\ and\
		\citenamefont {Belykh}}]{barabash2021partial}%
	\BibitemOpen
	\bibfield  {author} {\bibinfo {author} {\bibfnamefont {N.~V.}\ \bibnamefont
			{Barabash}}, \bibinfo {author} {\bibfnamefont {V.~N.}\ \bibnamefont
			{Belykh}}, \bibinfo {author} {\bibfnamefont {G.~V.}\ \bibnamefont {Osipov}},
		\ and\ \bibinfo {author} {\bibfnamefont {I.~V.}\ \bibnamefont {Belykh}},\
	}\href@noop {} {\bibfield  {journal} {\bibinfo  {journal} {Chaos}\ }\textbf
		{\bibinfo {volume} {31}} (\bibinfo {year} {2021})}\BibitemShut {NoStop}%
	\bibitem [{\citenamefont {G{\'o}mez-Gardenes}\ \emph
		{et~al.}(2011)\citenamefont {G{\'o}mez-Gardenes}, \citenamefont {G{\'o}mez},
		\citenamefont {Arenas},\ and\ \citenamefont {Moreno}}]{gomez2011explosive}%
	\BibitemOpen
	\bibfield  {author} {\bibinfo {author} {\bibfnamefont {J.}~\bibnamefont
			{G{\'o}mez-Gardenes}}, \bibinfo {author} {\bibfnamefont {S.}~\bibnamefont
			{G{\'o}mez}}, \bibinfo {author} {\bibfnamefont {A.}~\bibnamefont {Arenas}}, \
		and\ \bibinfo {author} {\bibfnamefont {Y.}~\bibnamefont {Moreno}},\
	}\href@noop {} {\bibfield  {journal} {\bibinfo  {journal} {Physical Review
				Letters}\ }\textbf {\bibinfo {volume} {106}},\ \bibinfo {pages} {128701}
		(\bibinfo {year} {2011})}\BibitemShut {NoStop}%
	\bibitem [{\citenamefont {Ji}\ \emph {et~al.}(2013)\citenamefont {Ji},
		\citenamefont {Peron}, \citenamefont {Menck}, \citenamefont {Rodrigues},\
		and\ \citenamefont {Kurths}}]{ji2013cluster}%
	\BibitemOpen
	\bibfield  {author} {\bibinfo {author} {\bibfnamefont {P.}~\bibnamefont
			{Ji}}, \bibinfo {author} {\bibfnamefont {T.~K.~D.}\ \bibnamefont {Peron}},
		\bibinfo {author} {\bibfnamefont {P.~J.}\ \bibnamefont {Menck}}, \bibinfo
		{author} {\bibfnamefont {F.~A.}\ \bibnamefont {Rodrigues}}, \ and\ \bibinfo
		{author} {\bibfnamefont {J.}~\bibnamefont {Kurths}},\ }\href@noop {}
	{\bibfield  {journal} {\bibinfo  {journal} {Physical Review Letters}\
		}\textbf {\bibinfo {volume} {110}},\ \bibinfo {pages} {218701} (\bibinfo
		{year} {2013})}\BibitemShut {NoStop}%
	\bibitem [{\citenamefont {Skardal}\ and\ \citenamefont
		{Arenas}(2014)}]{skardal2014disorder}%
	\BibitemOpen
	\bibfield  {author} {\bibinfo {author} {\bibfnamefont {P.~S.}\ \bibnamefont
			{Skardal}}\ and\ \bibinfo {author} {\bibfnamefont {A.}~\bibnamefont
			{Arenas}},\ }\href@noop {} {\bibfield  {journal} {\bibinfo  {journal}
			{Physical Review E}\ }\textbf {\bibinfo {volume} {89}},\ \bibinfo {pages}
		{062811} (\bibinfo {year} {2014})}\BibitemShut {NoStop}%
	\bibitem [{\citenamefont {Nishikawa}\ and\ \citenamefont
		{Motter}(2016)}]{nishikawa2016symmetric}%
	\BibitemOpen
	\bibfield  {author} {\bibinfo {author} {\bibfnamefont {T.}~\bibnamefont
			{Nishikawa}}\ and\ \bibinfo {author} {\bibfnamefont {A.~E.}\ \bibnamefont
			{Motter}},\ }\href@noop {} {\bibfield  {journal} {\bibinfo  {journal}
			{Physical Review Letters}\ }\textbf {\bibinfo {volume} {117}},\ \bibinfo
		{pages} {114101} (\bibinfo {year} {2016})}\BibitemShut {NoStop}%
	\bibitem [{\citenamefont {Nicolaou}\ \emph {et~al.}(2019)\citenamefont
		{Nicolaou}, \citenamefont {Eroglu},\ and\ \citenamefont
		{Motter}}]{nicolaou2019multifaceted}%
	\BibitemOpen
	\bibfield  {author} {\bibinfo {author} {\bibfnamefont {Z.~G.}\ \bibnamefont
			{Nicolaou}}, \bibinfo {author} {\bibfnamefont {D.}~\bibnamefont {Eroglu}}, \
		and\ \bibinfo {author} {\bibfnamefont {A.~E.}\ \bibnamefont {Motter}},\
	}\href@noop {} {\bibfield  {journal} {\bibinfo  {journal} {Physical Review
				X}\ }\textbf {\bibinfo {volume} {9}},\ \bibinfo {pages} {011017} (\bibinfo
		{year} {2019})}\BibitemShut {NoStop}%
	\bibitem [{\citenamefont {Kuramoto}\ and\ \citenamefont
		{Battogtokh}(2002)}]{kuramoto2002coexistence}%
	\BibitemOpen
	\bibfield  {author} {\bibinfo {author} {\bibfnamefont {Y.}~\bibnamefont
			{Kuramoto}}\ and\ \bibinfo {author} {\bibfnamefont {D.}~\bibnamefont
			{Battogtokh}},\ }\href@noop {} {\bibfield  {journal} {\bibinfo  {journal}
			{Nonlinear Phenomena in Complex Systems}\ }\textbf {\bibinfo {volume} {5}},\
		\bibinfo {pages} {380} (\bibinfo {year} {2002})}\BibitemShut {NoStop}%
	\bibitem [{\citenamefont {Abrams}\ and\ \citenamefont
		{Strogatz}(2004)}]{abrams2004chimera}%
	\BibitemOpen
	\bibfield  {author} {\bibinfo {author} {\bibfnamefont {D.~M.}\ \bibnamefont
			{Abrams}}\ and\ \bibinfo {author} {\bibfnamefont {S.~H.}\ \bibnamefont
			{Strogatz}},\ }\href@noop {} {\bibfield  {journal} {\bibinfo  {journal}
			{Physical Review Letters}\ }\textbf {\bibinfo {volume} {93}},\ \bibinfo
		{pages} {174102} (\bibinfo {year} {2004})}\BibitemShut {NoStop}%
	\bibitem [{\citenamefont {Abrams}\ \emph {et~al.}(2008)\citenamefont {Abrams},
		\citenamefont {Mirollo}, \citenamefont {Strogatz},\ and\ \citenamefont
		{Wiley}}]{abrams2008solvable}%
	\BibitemOpen
	\bibfield  {author} {\bibinfo {author} {\bibfnamefont {D.~M.}\ \bibnamefont
			{Abrams}}, \bibinfo {author} {\bibfnamefont {R.}~\bibnamefont {Mirollo}},
		\bibinfo {author} {\bibfnamefont {S.~H.}\ \bibnamefont {Strogatz}}, \ and\
		\bibinfo {author} {\bibfnamefont {D.~A.}\ \bibnamefont {Wiley}},\ }\href@noop
	{} {\bibfield  {journal} {\bibinfo  {journal} {Physical Review Letters}\
		}\textbf {\bibinfo {volume} {101}},\ \bibinfo {pages} {084103} (\bibinfo
		{year} {2008})}\BibitemShut {NoStop}%
	\bibitem [{\citenamefont {Panaggio}\ and\ \citenamefont
		{Abrams}(2015)}]{panaggio2015chimera}%
	\BibitemOpen
	\bibfield  {author} {\bibinfo {author} {\bibfnamefont {M.~J.}\ \bibnamefont
			{Panaggio}}\ and\ \bibinfo {author} {\bibfnamefont {D.~M.}\ \bibnamefont
			{Abrams}},\ }\href@noop {} {\bibfield  {journal} {\bibinfo  {journal}
			{Nonlinearity}\ }\textbf {\bibinfo {volume} {28}},\ \bibinfo {pages} {R67}
		(\bibinfo {year} {2015})}\BibitemShut {NoStop}%
	\bibitem [{\citenamefont {Zakharova}\ \emph {et~al.}(2014)\citenamefont
		{Zakharova}, \citenamefont {Kapeller},\ and\ \citenamefont
		{Sch{\"o}ll}}]{zakharova2014chimera}%
	\BibitemOpen
	\bibfield  {author} {\bibinfo {author} {\bibfnamefont {A.}~\bibnamefont
			{Zakharova}}, \bibinfo {author} {\bibfnamefont {M.}~\bibnamefont {Kapeller}},
		\ and\ \bibinfo {author} {\bibfnamefont {E.}~\bibnamefont {Sch{\"o}ll}},\
	}\href@noop {} {\bibfield  {journal} {\bibinfo  {journal} {Physical Review
				Letters}\ }\textbf {\bibinfo {volume} {112}},\ \bibinfo {pages} {154101}
		(\bibinfo {year} {2014})}\BibitemShut {NoStop}%
	\bibitem [{\citenamefont {Bolotov}\ \emph {et~al.}(2016)\citenamefont
		{Bolotov}, \citenamefont {Osipov},\ and\ \citenamefont
		{Pikovsky}}]{bolotov2016marginal}%
	\BibitemOpen
	\bibfield  {author} {\bibinfo {author} {\bibfnamefont {M.}~\bibnamefont
			{Bolotov}}, \bibinfo {author} {\bibfnamefont {G.}~\bibnamefont {Osipov}}, \
		and\ \bibinfo {author} {\bibfnamefont {A.}~\bibnamefont {Pikovsky}},\
	}\href@noop {} {\bibfield  {journal} {\bibinfo  {journal} {Physical Review
				E}\ }\textbf {\bibinfo {volume} {93}},\ \bibinfo {pages} {032202} (\bibinfo
		{year} {2016})}\BibitemShut {NoStop}%
	\bibitem [{\citenamefont {Bolotov}\ \emph {et~al.}(2018)\citenamefont
		{Bolotov}, \citenamefont {Smirnov}, \citenamefont {Osipov},\ and\
		\citenamefont {Pikovsky}}]{bolotov2018simple}%
	\BibitemOpen
	\bibfield  {author} {\bibinfo {author} {\bibfnamefont {M.}~\bibnamefont
			{Bolotov}}, \bibinfo {author} {\bibfnamefont {L.}~\bibnamefont {Smirnov}},
		\bibinfo {author} {\bibfnamefont {G.}~\bibnamefont {Osipov}}, \ and\ \bibinfo
		{author} {\bibfnamefont {A.}~\bibnamefont {Pikovsky}},\ }\href@noop {}
	{\bibfield  {journal} {\bibinfo  {journal} {Chaos: An Interdisciplinary
				Journal of Nonlinear Science}\ }\textbf {\bibinfo {volume} {28}},\ \bibinfo
		{pages} {045101} (\bibinfo {year} {2018})}\BibitemShut {NoStop}%
	\bibitem [{\citenamefont {Jaros}\ \emph {et~al.}(2015)\citenamefont {Jaros},
		\citenamefont {Maistrenko},\ and\ \citenamefont
		{Kapitaniak}}]{jaros2015chimera}%
	\BibitemOpen
	\bibfield  {author} {\bibinfo {author} {\bibfnamefont {P.}~\bibnamefont
			{Jaros}}, \bibinfo {author} {\bibfnamefont {Y.}~\bibnamefont {Maistrenko}}, \
		and\ \bibinfo {author} {\bibfnamefont {T.}~\bibnamefont {Kapitaniak}},\
	}\href@noop {} {\bibfield  {journal} {\bibinfo  {journal} {Physical Review
				E}\ }\textbf {\bibinfo {volume} {91}},\ \bibinfo {pages} {022907} (\bibinfo
		{year} {2015})}\BibitemShut {NoStop}%
	\bibitem [{\citenamefont {Maistrenko}\ \emph {et~al.}(2017)\citenamefont
		{Maistrenko}, \citenamefont {Brezetsky}, \citenamefont {Jaros}, \citenamefont
		{Levchenko},\ and\ \citenamefont {Kapitaniak}}]{maistrenko2017smallest}%
	\BibitemOpen
	\bibfield  {author} {\bibinfo {author} {\bibfnamefont {Y.}~\bibnamefont
			{Maistrenko}}, \bibinfo {author} {\bibfnamefont {S.}~\bibnamefont
			{Brezetsky}}, \bibinfo {author} {\bibfnamefont {P.}~\bibnamefont {Jaros}},
		\bibinfo {author} {\bibfnamefont {R.}~\bibnamefont {Levchenko}}, \ and\
		\bibinfo {author} {\bibfnamefont {T.}~\bibnamefont {Kapitaniak}},\
	}\href@noop {} {\bibfield  {journal} {\bibinfo  {journal} {Physical Review
				E}\ }\textbf {\bibinfo {volume} {95}},\ \bibinfo {pages} {010203} (\bibinfo
		{year} {2017})}\BibitemShut {NoStop}%
	\bibitem [{\citenamefont {Jaros}\ \emph {et~al.}(2018)\citenamefont {Jaros},
		\citenamefont {Brezetsky}, \citenamefont {Levchenko}, \citenamefont
		{Dudkowski}, \citenamefont {Kapitaniak},\ and\ \citenamefont
		{Maistrenko}}]{jaros2018solitary}%
	\BibitemOpen
	\bibfield  {author} {\bibinfo {author} {\bibfnamefont {P.}~\bibnamefont
			{Jaros}}, \bibinfo {author} {\bibfnamefont {S.}~\bibnamefont {Brezetsky}},
		\bibinfo {author} {\bibfnamefont {R.}~\bibnamefont {Levchenko}}, \bibinfo
		{author} {\bibfnamefont {D.}~\bibnamefont {Dudkowski}}, \bibinfo {author}
		{\bibfnamefont {T.}~\bibnamefont {Kapitaniak}}, \ and\ \bibinfo {author}
		{\bibfnamefont {Y.}~\bibnamefont {Maistrenko}},\ }\href@noop {} {\bibfield
		{journal} {\bibinfo  {journal} {Chaos: An Interdisciplinary Journal of
				Nonlinear Science}\ }\textbf {\bibinfo {volume} {28}},\ \bibinfo {pages}
		{011103} (\bibinfo {year} {2018})}\BibitemShut {NoStop}%
	\bibitem [{\citenamefont {Teichmann}\ and\ \citenamefont
		{Rosenblum}(2019)}]{teichmann2019solitary}%
	\BibitemOpen
	\bibfield  {author} {\bibinfo {author} {\bibfnamefont {E.}~\bibnamefont
			{Teichmann}}\ and\ \bibinfo {author} {\bibfnamefont {M.}~\bibnamefont
			{Rosenblum}},\ }\href@noop {} {\bibfield  {journal} {\bibinfo  {journal}
			{Chaos: An Interdisciplinary Journal of Nonlinear Science}\ }\textbf
		{\bibinfo {volume} {29}},\ \bibinfo {pages} {093124} (\bibinfo {year}
		{2019})}\BibitemShut {NoStop}%
	\bibitem [{\citenamefont {Munyayev}\ \emph {et~al.}(2022)\citenamefont
		{Munyayev}, \citenamefont {Bolotov}, \citenamefont {Smirnov}, \citenamefont
		{Osipov},\ and\ \citenamefont {Belykh}}]{munyayev2022stability}%
	\BibitemOpen
	\bibfield  {author} {\bibinfo {author} {\bibfnamefont {V.~O.}\ \bibnamefont
			{Munyayev}}, \bibinfo {author} {\bibfnamefont {M.~I.}\ \bibnamefont
			{Bolotov}}, \bibinfo {author} {\bibfnamefont {L.~A.}\ \bibnamefont
			{Smirnov}}, \bibinfo {author} {\bibfnamefont {G.~V.}\ \bibnamefont {Osipov}},
		\ and\ \bibinfo {author} {\bibfnamefont {I.~V.}\ \bibnamefont {Belykh}},\
	}\href@noop {} {\bibfield  {journal} {\bibinfo  {journal} {Physical Review
				E}\ }\textbf {\bibinfo {volume} {105}},\ \bibinfo {pages} {024203} (\bibinfo
		{year} {2022})}\BibitemShut {NoStop}%
	\bibitem [{\citenamefont {Belykh}\ \emph {et~al.}(2016)\citenamefont {Belykh},
		\citenamefont {Brister},\ and\ \citenamefont
		{Belykh}}]{belykh2016bistability}%
	\BibitemOpen
	\bibfield  {author} {\bibinfo {author} {\bibfnamefont {I.~V.}\ \bibnamefont
			{Belykh}}, \bibinfo {author} {\bibfnamefont {B.~N.}\ \bibnamefont {Brister}},
		\ and\ \bibinfo {author} {\bibfnamefont {V.~N.}\ \bibnamefont {Belykh}},\
	}\href@noop {} {\bibfield  {journal} {\bibinfo  {journal} {Chaos: An
				Interdisciplinary Journal of Nonlinear Science}\ }\textbf {\bibinfo {volume}
			{26}},\ \bibinfo {pages} {094822} (\bibinfo {year} {2016})}\BibitemShut
	{NoStop}%
	\bibitem [{\citenamefont {Brister}\ \emph {et~al.}(2020)\citenamefont
		{Brister}, \citenamefont {Belykh},\ and\ \citenamefont
		{Belykh}}]{brister2020three}%
	\BibitemOpen
	\bibfield  {author} {\bibinfo {author} {\bibfnamefont {B.~N.}\ \bibnamefont
			{Brister}}, \bibinfo {author} {\bibfnamefont {V.~N.}\ \bibnamefont {Belykh}},
		\ and\ \bibinfo {author} {\bibfnamefont {I.~V.}\ \bibnamefont {Belykh}},\
	}\href@noop {} {\bibfield  {journal} {\bibinfo  {journal} {Physical Review
				E}\ }\textbf {\bibinfo {volume} {101}},\ \bibinfo {pages} {062206} (\bibinfo
		{year} {2020})}\BibitemShut {NoStop}%
	\bibitem [{\citenamefont {Ronge}\ and\ \citenamefont
		{Zaks}(2021)}]{ronge2021splay}%
	\BibitemOpen
	\bibfield  {author} {\bibinfo {author} {\bibfnamefont {R.}~\bibnamefont
			{Ronge}}\ and\ \bibinfo {author} {\bibfnamefont {M.~A.}\ \bibnamefont
			{Zaks}},\ }\href@noop {} {\bibfield  {journal} {\bibinfo  {journal} {The
				European Physical Journal Special Topics}\ }\textbf {\bibinfo {volume}
			{230}},\ \bibinfo {pages} {2717} (\bibinfo {year} {2021})}\BibitemShut
	{NoStop}%
	\bibitem [{\citenamefont {Berner}\ \emph {et~al.}(2021)\citenamefont {Berner},
		\citenamefont {Yanchuk}, \citenamefont {Maistrenko},\ and\ \citenamefont
		{Scholl}}]{berner2021generalized}%
	\BibitemOpen
	\bibfield  {author} {\bibinfo {author} {\bibfnamefont {R.}~\bibnamefont
			{Berner}}, \bibinfo {author} {\bibfnamefont {S.}~\bibnamefont {Yanchuk}},
		\bibinfo {author} {\bibfnamefont {Y.}~\bibnamefont {Maistrenko}}, \ and\
		\bibinfo {author} {\bibfnamefont {E.}~\bibnamefont {Scholl}},\ }\href@noop {}
	{\bibfield  {journal} {\bibinfo  {journal} {Chaos: An Interdisciplinary
				Journal of Nonlinear Science}\ }\textbf {\bibinfo {volume} {31}},\ \bibinfo
		{pages} {073128} (\bibinfo {year} {2021})}\BibitemShut {NoStop}%
	\bibitem [{\citenamefont {Tsimring}\ \emph {et~al.}(2005)\citenamefont
		{Tsimring}, \citenamefont {Rulkov}, \citenamefont {Larsen},\ and\
		\citenamefont {Gabbay}}]{tsimring2005repulsive}%
	\BibitemOpen
	\bibfield  {author} {\bibinfo {author} {\bibfnamefont {L.}~\bibnamefont
			{Tsimring}}, \bibinfo {author} {\bibfnamefont {N.}~\bibnamefont {Rulkov}},
		\bibinfo {author} {\bibfnamefont {M.}~\bibnamefont {Larsen}}, \ and\ \bibinfo
		{author} {\bibfnamefont {M.}~\bibnamefont {Gabbay}},\ }\href@noop {}
	{\bibfield  {journal} {\bibinfo  {journal} {Physical Review Letters}\
		}\textbf {\bibinfo {volume} {95}},\ \bibinfo {pages} {014101} (\bibinfo
		{year} {2005})}\BibitemShut {NoStop}%
	\bibitem [{\citenamefont {Gao}\ \emph {et~al.}(2019)\citenamefont {Gao},
		\citenamefont {Fu}, \citenamefont {Cai}, \citenamefont {Yang},\ and\
		\citenamefont {Eugene~Stanley}}]{gao2019repulsive}%
	\BibitemOpen
	\bibfield  {author} {\bibinfo {author} {\bibfnamefont {Y.-C.}\ \bibnamefont
			{Gao}}, \bibinfo {author} {\bibfnamefont {C.-J.}\ \bibnamefont {Fu}},
		\bibinfo {author} {\bibfnamefont {S.-M.}\ \bibnamefont {Cai}}, \bibinfo
		{author} {\bibfnamefont {C.}~\bibnamefont {Yang}}, \ and\ \bibinfo {author}
		{\bibfnamefont {H.}~\bibnamefont {Eugene~Stanley}},\ }\href@noop {}
	{\bibfield  {journal} {\bibinfo  {journal} {Chaos: An Interdisciplinary
				Journal of Nonlinear Science}\ }\textbf {\bibinfo {volume} {29}},\ \bibinfo
		{pages} {053130} (\bibinfo {year} {2019})}\BibitemShut {NoStop}%
	\bibitem [{\citenamefont {Belykh}\ and\ \citenamefont
		{Shilnikov}(2008)}]{belykh2008weak}%
	\BibitemOpen
	\bibfield  {author} {\bibinfo {author} {\bibfnamefont {I.}~\bibnamefont
			{Belykh}}\ and\ \bibinfo {author} {\bibfnamefont {A.}~\bibnamefont
			{Shilnikov}},\ }\href@noop {} {\bibfield  {journal} {\bibinfo  {journal}
			{Physical Review Letters}\ }\textbf {\bibinfo {volume} {101}},\ \bibinfo
		{pages} {078102} (\bibinfo {year} {2008})}\BibitemShut {NoStop}%
	\bibitem [{\citenamefont {Nishikawa}\ and\ \citenamefont
		{Motter}(2010)}]{nishikawa2010network}%
	\BibitemOpen
	\bibfield  {author} {\bibinfo {author} {\bibfnamefont {T.}~\bibnamefont
			{Nishikawa}}\ and\ \bibinfo {author} {\bibfnamefont {A.~E.}\ \bibnamefont
			{Motter}},\ }\href@noop {} {\bibfield  {journal} {\bibinfo  {journal}
			{Proceedings of the National Academy of Sciences}\ }\textbf {\bibinfo
			{volume} {107}},\ \bibinfo {pages} {10342} (\bibinfo {year}
		{2010})}\BibitemShut {NoStop}%
	\bibitem [{\citenamefont {Belykh}\ \emph {et~al.}(2015)\citenamefont {Belykh},
		\citenamefont {Reimbayev},\ and\ \citenamefont
		{Zhao}}]{belykh2015synergistic}%
	\BibitemOpen
	\bibfield  {author} {\bibinfo {author} {\bibfnamefont {I.}~\bibnamefont
			{Belykh}}, \bibinfo {author} {\bibfnamefont {R.}~\bibnamefont {Reimbayev}}, \
		and\ \bibinfo {author} {\bibfnamefont {K.}~\bibnamefont {Zhao}},\ }\href@noop
	{} {\bibfield  {journal} {\bibinfo  {journal} {Physical Review E}\ }\textbf
		{\bibinfo {volume} {91}},\ \bibinfo {pages} {062919} (\bibinfo {year}
		{2015})}\BibitemShut {NoStop}%
	\bibitem [{\citenamefont {Reimbayev}\ \emph {et~al.}(2017)\citenamefont
		{Reimbayev}, \citenamefont {Daley},\ and\ \citenamefont
		{Belykh}}]{reimbayev2017two}%
	\BibitemOpen
	\bibfield  {author} {\bibinfo {author} {\bibfnamefont {R.}~\bibnamefont
			{Reimbayev}}, \bibinfo {author} {\bibfnamefont {K.}~\bibnamefont {Daley}}, \
		and\ \bibinfo {author} {\bibfnamefont {I.}~\bibnamefont {Belykh}},\
	}\href@noop {} {\bibfield  {journal} {\bibinfo  {journal} {Philosophical
				Transactions of the Royal Society A: Mathematical, Physical and Engineering
				Sciences}\ }\textbf {\bibinfo {volume} {375}},\ \bibinfo {pages} {20160282}
		(\bibinfo {year} {2017})}\BibitemShut {NoStop}%
	\bibitem [{\citenamefont {Delabays}(2019)}]{delabays2019dynamical}%
	\BibitemOpen
	\bibfield  {author} {\bibinfo {author} {\bibfnamefont {R.}~\bibnamefont
			{Delabays}},\ }\href@noop {} {\bibfield  {journal} {\bibinfo  {journal}
			{Chaos: An Interdisciplinary Journal of Nonlinear Science}\ }\textbf
		{\bibinfo {volume} {29}},\ \bibinfo {pages} {113129} (\bibinfo {year}
		{2019})}\BibitemShut {NoStop}%
	\bibitem [{\citenamefont {Seliger}\ \emph {et~al.}(2002)\citenamefont
		{Seliger}, \citenamefont {Young},\ and\ \citenamefont
		{Tsimring}}]{seliger2002plasticity}%
	\BibitemOpen
	\bibfield  {author} {\bibinfo {author} {\bibfnamefont {P.}~\bibnamefont
			{Seliger}}, \bibinfo {author} {\bibfnamefont {S.~C.}\ \bibnamefont {Young}},
		\ and\ \bibinfo {author} {\bibfnamefont {L.~S.}\ \bibnamefont {Tsimring}},\
	}\href@noop {} {\bibfield  {journal} {\bibinfo  {journal} {Physical Review
				E}\ }\textbf {\bibinfo {volume} {65}},\ \bibinfo {pages} {041906} (\bibinfo
		{year} {2002})}\BibitemShut {NoStop}%
	\bibitem [{\citenamefont {Niyogi}\ and\ \citenamefont
		{English}(2009)}]{niyogi2009learning}%
	\BibitemOpen
	\bibfield  {author} {\bibinfo {author} {\bibfnamefont {R.~K.}\ \bibnamefont
			{Niyogi}}\ and\ \bibinfo {author} {\bibfnamefont {L.~Q.}\ \bibnamefont
			{English}},\ }\href@noop {} {\bibfield  {journal} {\bibinfo  {journal}
			{Physical Review E}\ }\textbf {\bibinfo {volume} {80}},\ \bibinfo {pages}
		{066213} (\bibinfo {year} {2009})}\BibitemShut {NoStop}%
	\bibitem [{\citenamefont {Kiss}\ \emph {et~al.}(2005)\citenamefont {Kiss},
		\citenamefont {Zhai},\ and\ \citenamefont {Hudson}}]{kiss2005predicting}%
	\BibitemOpen
	\bibfield  {author} {\bibinfo {author} {\bibfnamefont {I.~Z.}\ \bibnamefont
			{Kiss}}, \bibinfo {author} {\bibfnamefont {Y.}~\bibnamefont {Zhai}}, \ and\
		\bibinfo {author} {\bibfnamefont {J.~L.}\ \bibnamefont {Hudson}},\
	}\href@noop {} {\bibfield  {journal} {\bibinfo  {journal} {Physical Review
				Letters}\ }\textbf {\bibinfo {volume} {94}},\ \bibinfo {pages} {248301}
		(\bibinfo {year} {2005})}\BibitemShut {NoStop}%
	\bibitem [{\citenamefont {Goldobin}\ \emph {et~al.}(2013)\citenamefont
		{Goldobin}, \citenamefont {Kleiner}, \citenamefont {Koelle},\ and\
		\citenamefont {Mints}}]{goldobin2013phase}%
	\BibitemOpen
	\bibfield  {author} {\bibinfo {author} {\bibfnamefont {E.}~\bibnamefont
			{Goldobin}}, \bibinfo {author} {\bibfnamefont {R.}~\bibnamefont {Kleiner}},
		\bibinfo {author} {\bibfnamefont {D.}~\bibnamefont {Koelle}}, \ and\ \bibinfo
		{author} {\bibfnamefont {R.}~\bibnamefont {Mints}},\ }\href@noop {}
	{\bibfield  {journal} {\bibinfo  {journal} {Physical Review Letters}\
		}\textbf {\bibinfo {volume} {111}},\ \bibinfo {pages} {057004} (\bibinfo
		{year} {2013})}\BibitemShut {NoStop}%
	\bibitem [{\citenamefont {Komarov}\ and\ \citenamefont
		{Pikovsky}(2013)}]{komarov2013multiplicity}%
	\BibitemOpen
	\bibfield  {author} {\bibinfo {author} {\bibfnamefont {M.}~\bibnamefont
			{Komarov}}\ and\ \bibinfo {author} {\bibfnamefont {A.}~\bibnamefont
			{Pikovsky}},\ }\href@noop {} {\bibfield  {journal} {\bibinfo  {journal}
			{Physical Review Letters}\ }\textbf {\bibinfo {volume} {111}},\ \bibinfo
		{pages} {204101} (\bibinfo {year} {2013})}\BibitemShut {NoStop}%
	\bibitem [{\citenamefont {Skardal}\ \emph {et~al.}(2011)\citenamefont
		{Skardal}, \citenamefont {Ott},\ and\ \citenamefont
		{Restrepo}}]{skardal2011cluster}%
	\BibitemOpen
	\bibfield  {author} {\bibinfo {author} {\bibfnamefont {P.~S.}\ \bibnamefont
			{Skardal}}, \bibinfo {author} {\bibfnamefont {E.}~\bibnamefont {Ott}}, \ and\
		\bibinfo {author} {\bibfnamefont {J.~G.}\ \bibnamefont {Restrepo}},\
	}\href@noop {} {\bibfield  {journal} {\bibinfo  {journal} {Physical Review
				E}\ }\textbf {\bibinfo {volume} {84}},\ \bibinfo {pages} {036208} (\bibinfo
		{year} {2011})}\BibitemShut {NoStop}%
	\bibitem [{\citenamefont {Sakaguchi}(2006)}]{sakaguchi2006instability}%
	\BibitemOpen
	\bibfield  {author} {\bibinfo {author} {\bibfnamefont {H.}~\bibnamefont
			{Sakaguchi}},\ }\href@noop {} {\bibfield  {journal} {\bibinfo  {journal}
			{Physical Review E}\ }\textbf {\bibinfo {volume} {73}},\ \bibinfo {pages}
		{031907} (\bibinfo {year} {2006})}\BibitemShut {NoStop}%
\end{thebibliography}
\end{document}